\documentclass[preprint,preprintnumbers,showkeys,amsmath,amssymb,pra]{revtex4}% Physical Review B

\usepackage{epsf}
\usepackage{graphicx}% Include figure files
\usepackage{dcolumn}% Align table columns on decimal point
\usepackage{bm}% bold math
\usepackage{color}
\everymath{\displaystyle}
\begin{document}

\title{Field of a moving locked charge as applied to beam-beam interactions in storage rings}

\author{Alexander J. Silenko}

\affiliation{Research Institute for
Nuclear Problems, Belarusian State University, Minsk 220030, Belarus,\\
Bogoliubov Laboratory of Theoretical Physics, Joint Institute for Nuclear Research,
Dubna 141980, Russia}  %%%%%%%%%%%%%%%%%%%%%%%%%%% second institution!!!

\date{\today}
\begin {abstract}
It is shown that the Lorentz transformation cannot in general be
formally applied to potentials and fields of particles locked in a
certain region. In particular, this property relates to nucleons
in nuclei and to particles and nuclei in storage rings. Even if
they
%nucleons in a nucleus and particles and nuclei in storage rings
move with high velocities, their electric fields are defined by
the Coulomb law. The result obtained is rather important for the
planned deuteron electric-dipole-moment experiment in storage
rings.
\end{abstract}

\maketitle

\section{Introduction}

A simultaneous use of two beams in electric-dipole-moment (EDM)
experiments may be helpful. Injecting the beam clockwise and
counter-clockwise allows one to cancel some systematical errors.
However, the beam dynamics and the spin motion are affected by the
beam-beam interaction. In particular, spins of particles moving
clockwise rotate about the radial axis in the electric and
magnetic fields of the counter-clockwise beam. To describe the
beam dynamics and the spin motion in such an experiment, beam-beam
interactions should be correctly described.

At first sight, this task seems to be very simple because a
distribution of charges and currents is well-defined. The scalar
and vector potentials of moving particles can be determined with
the use of Lorentz transformations. However, the problem is rather
nontrivial. We show that interactions with particles locked in a
certain region have some important peculiarities.

In storage rings, particles and nuclei are locked. This
circumstance influences interactions between two beams. % in a storage ring.
We consider how this effect manifests in a particle
spin motion and ascertain its relation to spin dynamics in
electric-dipole-moment (EDM) experiments.

We use the system of units $\hbar=1,~c=1$. We include $\hbar$ and
$c$ into some equations when this inclusion clarifies the problem.

\section{Electric field of protons moving in a nucleus}\label{nucleus}

It is instructive to consider the electric field of protons moving
in a nucleus. It is well known that the scalar potential and the
strength of the field of a nucleus at rest are equal to
\begin{eqnarray} \Phi=\frac{Ze}{r}, ~~~ \bm E=\frac{Ze\bm r}{r^3}.
\label{eqC}\end{eqnarray} Numerous experiments have shown that Eq.
(\ref{eqC}) is valid and the measured number $Z$ is integer.
Therefore, motion of charged nucleons (protons) does not affect
the potential and the electric field of the nucleus.

However, this result is nontrivial. Protons and neutrons move in
the nucleus with high velocities. The simplest way for a description of
the electric field of the nucleus is an assumption that electric
fields of all protons are independent and may be summarized. Let the
nucleus be at rest in the lab frame. In this frame, the scalar potential of the moving
charge $e$ can be obtained with the Lorentz transformation from its
rest frame:
\begin{eqnarray}\Phi=\frac{\Phi_0}{\sqrt{1-\bm\beta^2}}=\gamma\Phi_0=\frac{\gamma
    e}{r_0},~~~\bm\beta=\frac{\bm v}{c},\label{eqPhi}\end{eqnarray} where
$\gamma$ is the Lorentz factor.
Evidently, $r_0=r$ when the vectors $\bm v$ and $\bm r$ are orthogonal
and $r_0=\gamma r$ when these vectors are collinear. Thus, $\Phi=e/r$
only in the latter case and $\Phi$ is greater than $e/r$ in other
cases. In particular, $\Phi=\Phi_0(1+\bm\beta^2/2)$ in the nonrelativistic approximation. The difference between $\Phi$ and $e/r$ can be
 easily checked. While the nucleons in the nucleus move in
 random directions and an average velocity of each nucleon is equal to zero,
 the resulting potential of the nucleus differs from the potential (\ref{eqC}).

The electric field corresponding to Eq. (\ref{eqPhi}) is equal to
\begin{eqnarray}
\bm E=\left[\bm r-\frac{\gamma}{\gamma+1}\bm\beta(\bm\beta\cdot\bm r)\right]\frac{\gamma e}{r^3}.
\label{eqE}\end{eqnarray}

Averaging with respect to a proton motion in the nucleus results in
\begin{eqnarray}
\overline{\bm E}=\frac{(2\gamma+1) e\bm r}{3r^3}.
\label{eqEvd}\end{eqnarray} 
This result also disagrees with the conventional electric field of
nuclei given by Eq. (\ref{eqC}) and with experimental data confirming
this equation. The discrepancy can be reinforced with taking into
account quark structure of nucleons because velocities of quarks are
much closer to the light velocity as those of nucleons. Numerous
experiments performed in nuclear and atomic physics explicitly
demonstrate that we need not consider the internal structure of the
nuclei.

The average potential of the nucleon is given by
$$ \overline{\Phi}=\int{\overline{\bm E}\cdot d\bm
    r}=\frac{(2\gamma+1) e}{3r}$$
and the average potential of the nucleus is equal
  to
\begin{eqnarray} \overline{\Phi}_N=\frac{e}{3}\sum_{i=1}^{Z}{\frac{2\gamma_i+1}{r_i}}.\label{eqPhiii}\end{eqnarray}
Evidently, it significantly differs from the conventional potential (\ref{eqC}).

Reasons of this situation can be
 explained. We can show that standard Lorentz transformations are inapplicable
to locked particles. Let us consider an interaction of a slowly
moving electron with an electric field of a nucleus. The kinetic
energy of the electron slowly increases accordingly to a decrease
of its potential energy.
%Let the distance between the electron and nucleus be much bigger as compared with the nucleus radius.
Protons in the nucleus move much more quickly than the electron.
Therefore, a change of the electron position during a proton
oscillation can be neglected. This means that we may consider the
proton moving in the electrostatic field of the electron. The
potential and kinetic energies of the proton change due to the
interaction with the electron field. The field of the slow
electron is characterized by the potential $-|e|/r$ and by the
strength $-|e|\bm r/r^3$. Therefore, the potential energy of an
electron-proton interaction is always equal to $-e^2/r$. The
interaction energy of any proton is defined by this expression
while the quantity $r$ can differ for different protons. As a
result, the total energy of the electron-nucleus interaction is
equal to $-e^2\sum_{i=1}^{Z}{1/r_i}$. Due to a symmetric arrangement of protons in the nucleus, its potential takes the form (\ref{eqC}).

The electric field of a \emph{moving} nucleus can be obtained with
an appropriate Lorentz transformation and is given by Eq.
(\ref{eqE}) with the substitution of $Ze$ for $e$. In this case, $\bm\beta$ characterizes the velocity of the
center of mass of the nucleus.

The results of this discussion relate
  not only to nuclei with $Z>1$ but also to the deuteron with
  $Z=1$. Their importance for the
deuteron is caused by a motion of the proton about the center of mass of the
two nucleus. The difference between the fields of free and locked
particles relates also to proton beams in closed spaces like storage rings.
This problem is considered in the next section.

\section{Beam-beam interactions in storage rings}

We can now consider the beam-beam interactions in storage rings. While
the spherical symmetry is characteristic for nuclei, the cylindrical
symmetry and the cylindrical coordinates are characteristic for
storage rings.
We analyze the case when two rings are simultaneously used.
Charged particles forming the two beams are \emph{locked} in the storage
rings like protons in the nucleus. Thus, the scalar potential and
the electric field strength of any particle in the \emph{lab
frame} are equal to $e/r$ and $-e\bm r/r^3$, respectively, where
$r$ is also defined in the lab frame. An integration of the total
electric field strength of the beam allows one to calculate the
total potential of this beam. This means that the energy of the
beam-beam interactions in storage rings does not depend on the
momenta of the beams and is defined by the Coulomb potential.

The electric field of a particle beam is defined by the electric field
of a ring of charge \cite{ringofcharge}. Near the particle beam (when
the distance to the beam is much less than the ring radius) this field
is almost orthogonal to the tangent to the ring and has a shape of a
squeezed torus. The field strength
can be approximated by
that of an infinite charged filament formed by the particles and is given by
\begin{eqnarray}
\bm E=E\frac{\bm \rho}{\rho}, ~~~ E=\frac{2\tau}{\rho},
\label{eqfil}\end{eqnarray} where $\rho$ is the distance to the filament and $\tau$ is the charge of the unit of its length.

Otherwise, a formal use of Lorentz transformed potentials brings a very different result. In this case, the electric field of the infinite charged filament formed by moving particles is equal to
\begin{eqnarray}
E=\frac{2\tau\gamma}{\rho}. \label{eqfilwn}\end{eqnarray} A
difference between Eqs. (\ref{eqfil}) and (\ref{eqfilwn}) is
rather significant.

We should add that the motion of particles
creates also an electric current which magnetic field is given by
\begin{eqnarray}
\bm B=\frac{2\tau\bm\beta\times\bm \rho}{\rho^2}.
\label{eqfilwm}\end{eqnarray}

The problem directly relates to EDM experiments in storage rings.
In these experiments, one can use two beams consisting from
particles moving in the clockwise and counterclockwise directions
with the same momentum \cite{dEDM,pEDM}. When the two beams are
separated in space \cite{dEDM}, particles of one beam create a
vertical electric field and a radial magnetic one acting on
particles of another beam. The most important effect of beam-beam
interactions is an action of these fields on the spin motion. The
angular velocity of spin motion is defined by the
Thomas-Bargmann-Michel-Telegdi equation \cite{T-BMT} extended on
the EDM \cite{EDM}:
\begin{equation} \begin{array} {c}
\bm \Omega=-\frac{e}{mc}\left[\left(G+\frac{1}{\gamma}\right){\bm B}-\frac{\gamma G}{\gamma+1}({\bm\beta}\cdot{\bm B}){\bm\beta}-\left(G+\frac{1}{\gamma+1}\right)\bm\beta\times{\bm E}\right.\\
+\left.\frac{\eta}{2}\left({\bm E}-\frac{\gamma}{\gamma+1}(\bm\beta\cdot{\bm E})\bm\beta+\bm\beta\times {\bm B}\right)\right].
\end{array} \label{Nelsonh} \end{equation}
Here $G=(g-2)/2,~g=2mc\mu/(e\hbar s),~\eta=2mcd/(e\hbar s)$, $s$
is the spin quantum number, and $\mu$ and $d$ are the magnetic and
electric dipole moments. The effects of the vertical magnetic
field and the radial electric one on the EDM consist in a spin
rotation about the radial axis. Equation (\ref{Nelsonh}) shows
that such a rotation can also be conditioned by the vertical
electric field and the radial magnetic one acting on the magnetic
dipole moment. Therefore, the two latter fields may cause the spin
rotation imitating the presence of the EDM.

We should take into account that the average vertical force acting
on the beam is zero. When magnetic focusing is used, the resulting
radial magnetic field is equal to the sum of the magnetic fields
created by the motion of particles of another beam [see Eq.
(\ref{eqfilwm})] and by focusing magnets. The resulting radial
magnetic field causes the force which should be equal on the
average to the force originated in the vertical electric field.
The latter field is defined by Eq. (\ref{eqfil}) [or,
alternatively, by Eq. (\ref{eqfilwn})]. When one uses electric
focusing, the average force caused by the summary electric field
of the beam and of the focusing electric quadrupoles
counterbalances the force conditioned by the magnetic field
(\ref{eqfilwm}). The spin turn about the radial axis takes place
in the both cases.

\section{Spin motion caused by the beam-beam interactions in the deuteron EDM experiment}

One plans to use two beams circulating clockwise and
counterclockwise in the deuteron \cite{dEDM} and proton
\cite{pEDM} EDM experiments in storage rings.
%
%
%
%
%In the proton EDM
%experiment, the clockwise and counterclockwise beams are joined
%(the split between them is of the order of a picometer
%\cite{pEDM}). As a result, the beam-beam interactions are not
%important. In the deuteron EDM experiment, the situation is
%different because one plans to have two independent rings located
%on top of each other, about 40 cm apart \cite{dEDM}. The beams in the two rings will move clockwise and counterclockwise.
%
Therefore, it is important to calculate an effect of electric field of one beam on the particle spin motion in another beam.
%
%
% while
%this field %effect
%can in principle be eliminated by shielding.
%
%
%
To describe the spin
motion in storage ring EDM experiments, it is convenient to
present Eq. (\ref{Nelsonh}) in cylindrical coordinates
\cite{PhysRevSpTopAB}:
\begin{equation} \begin{array} {c}
\bm \Omega'=-\frac{e}{mc}\left\{G{\bm B}-\frac{\gamma
    G}{\gamma+1}({\bm\beta}\cdot{\bm
    B})\bm\beta+\left(\frac{1}{\gamma^2-1}-
G\right)(\bm\beta\times{\bm E})+\frac{1}{\gamma}\left[\bm B_\parallel-\frac{1}{\beta^2}(\bm\beta\times{\bm E})_\parallel\right]\right.\\
+\left.\frac{\eta}{2}\left({\bm E}-\frac{\gamma}{\gamma+1}(\bm\beta\cdot{\bm E})\bm\beta+\bm\beta\times {\bm B}\right)\right\}.
\end{array} \label{PRSpTAcBeams} \end{equation}
The sign $\parallel$ means a horizontal
projection for any vector.

Equation (\ref{PRSpTAcBeams}) shows that the special connection
between the vertical magnetic field and the radial electric one,
$$E_r=-\frac{G\beta\gamma^2 B_z}{1-G\beta^2\gamma^2},$$
cancels the spin rotation about the vertical axis conditioned by the magnetic moment. However, there is the spin rotation about the radial axis. This rotation is caused by the EDM and is given by
\begin{equation} \begin{array} {c}
{\bm \Omega'}_{EDM}=\frac{e\eta}{2mc}\cdot\frac{\beta B_z}{1-G\beta^2\gamma^2}\bm e_r=\frac{d}{\hbar}\cdot\frac{\beta B_z}{1-G\beta^2\gamma^2}\bm e_r.
\end{array} \label{EDMr} \end{equation}

The vertical electric field is counterbalanced by the %focusing
radial magnetic field $B_r=-E_z/\beta$. Cumbersome but simple
calculations define the overall effect of these two fields on the
spin motion:
\begin{equation} \begin{array} {c}
{\bm \Omega'}_{b-b}=\frac{e(1+G)}{mc\beta\gamma^2}E_z\bm e_r.
\end{array} \label{bb} \end{equation}
These calculations are based on Eq. (\ref{eqfil}). The use of Eq.
(\ref{eqfilwn}) increases the angular velocity of the spin
rotation by the factor of $\gamma$.

In the deuteron EDM experiment, one plans to have two independent rings located
on top of each other, about 40 cm apart \cite{dEDM}.
Each ring will contain $1\times10^{11}$
nuclei rotating clockwise in one ring and counterclockwise in another one. Since the circumferences of the rings are 82.955 m
\cite{dEDM}, the vertical electric field is 8.7 V. For a ring with
the beam momentum $p=1$ GeV/c and the vertical magnetic field
$B_z=0.5$ T \cite{dEDM}, a comparison of Eqs. (\ref{EDMr}) and
(\ref{bb}) shows that the spin rotation defined by Eqs. (\ref{eqfil}) and (\ref{bb}) corresponds to the
deuteron EDM $d=1.9\times10^{-21}$ $e\cdot$cm. The effect of the
electric field conditioned by the moving beam is therefore rather
important and the use of the right equation (\ref{eqfil}) is
necessary. The large value of the systematical correction for the
beam-beam interactions (i.e., for the vertical electric field
originated from another beam) as compared with the planned
experimental sensitivity of $d=1\times10^{-29}$ $e\cdot$cm proves
a necessity of shielding this field.

Shielding the influence of the electric field of one beam on another beam is stipulated in the deuteron EDM experiment
\cite{dEDM}. When the two rings are made of a conductive material, they shield the electric and magnetic fields of the stored beams. In this case, one needs to take into account a beam-wall interaction caused by a mirror current running on the walls of the rings.

The situation is very different in the proton EDM experiment. In this
experiment, the clockwise and counterclockwise beams are joined
(the split between them is of the order of a picometer
\cite{pEDM}). Therefore, we should consider a particle motion and a spin rotation in the resulting fields created by the two beams. The magnetic field is almost zero in the lab frame due to a clockwise and counterclockwise motion of particles. The electric field acting on a particle is the field of a charged cylinder. The cylinder radius $\rho_0$ is equal to the beam radius. In the considered case, the electric field is given by 
\begin{eqnarray}
\bm E=2\pi\eta\bm \rho, ~~~ \rho\leq\rho_0,\label{eqfillm}\end{eqnarray}\begin{eqnarray}
\bm E=\frac{2\pi\eta\rho_0^2\bm \rho}{\rho^2}, ~~~ \rho>\rho_0,
\label{eqfilll}\end{eqnarray}
where $\eta$ is the charge density. As compared with Eq. (\ref{eqfil}), $\tau=\pi\rho_0^2\eta$.

Equations (\ref{eqfillm}),(\ref{eqfilll}) define the fields acting in both the vertical and radial directions ($E_z=E\cos{\theta},~E_r=E\sin{\theta}$). It is easy to show that the action of these vertical and radial fields consists in a change of frequencies of the vertical and radial betatron oscillations caused by focusing fields. Equations (\ref{eqfillm}),(\ref{eqfilll}) demonstrate that the average electric field of the two beams acting on the spin is almost zero. Its negligible nonzero value is conditioned by the small asymmetry of beam positions.

We should underline that Eq. (\ref{eqfil}) defines the electric
field which acts not only on another beam but also on ring
magnets, electric plates and so on. The electric field in an
inertial frame \emph{moving relatively the storage ring} can be
obtained with the corresponding Lorentz transformation (cf. the
end of Sec. \ref{nucleus}).

\section{Summary}

We have ascertained that the Lorentz transformation cannot in
general be formally applied to potentials and fields of particles
locked in a certain region. In particular, this property relates
to nucleons in nuclei and to particles and nuclei in storage
rings. Even if they
%nucleons in a nucleus and particles and nuclei in storage rings
move with high velocities, their electric fields
are defined by the Coulomb law. The result obtained is rather
important for the planned deuteron EDM experiment in storage
rings.

%The same situation take place with a calculation of the vector potential of the beams.

\section*{Acknowledgements}

The author is grateful to Y.K. Semertzidis for useful discussions.
The work was supported in part by the Belarusian Republican
Foundation for Fundamental Research (Grant No. $\Phi$14D-007) and
by the Heisenberg-Landau program of the German Ministry for
Science and Technology (Bundesministerium f\"{u}r Bildung und
Forschung).

%\footnotesize
%\renewcommand{\baselinestretch}{1.4}

\end{document}